\documentclass{aa}
\usepackage{graphics}
\def\boi{B\,{\sc i}}
\def\fei{Fe\,{\sc i}}
\def\coi{Co\,{\sc i}}
\def\teff{${T_{\rm eff}}$}
\def\logg{$\log g$}
\def\gf{{\it gf}}
\def\eps{$ \epsilon $}
\def\etal{et~al.}
\def\bd-13{BD~$-$13~3442}

\begin{document}

\thesaurus{7(08.09.2 \bd-13; 08.01.1; 08.01.3; 09.03.2; 10.01.1; 
10.05.1)}
 
\title{
Boron in the very metal-poor star \bd-13\thanks{This
research was based on observations obtained with the NASA/ESA {\it
Hubble Space Telescope} through the Space Telescope Science Institute,
which is operated by the Association of Universities for Research in
Astronomy, Inc., under NASA contract NAS5-26555.}}
 
\author{D.K. Duncan\inst{1,2} \and L.M. Rebull\inst{1} \and F. Primas\inst{1,3} \and A.M. Boesgaard\inst{4} \and Constantine P. Deliyannis\inst{5,6}
\and L.M. Hobbs\inst{7} \and J.R. King\inst{8} \and S.G. Ryan\inst{9}}
\institute{University of Chicago, Department of Astronomy and Astrophysics,
       5640 S. Ellis Ave, Chicago, IL 60637 USA 
        (duncan, rebull, primas@oddjob.uchicago.edu) \and
Adler Planetarium, Chicago, IL, USA \and
European Southern Observatory, Karl-Schwarzschild Str. 2, 
       D-85748 Garching b. M\"{u}nich \and
University of Hawaii Institute for Astronomy, 2680 Woodlawn Dr, 
       Honolulu, HI 96822 USA (boes@galileo.ifa.hawaii.edu) \and
Beatrice Watson Parrent Postdoctoral Fellow; Hubble Postdoctoral Fellow \and
Yale University, Department of Astronomy, P.O. Box 208101, New 
        Haven CT 06520-8101 USA (con@athena.astro.yale.edu) \and
University of Chicago, Yerkes Observatory, Williams Bay, WI 
        53191-0258 USA (hobbs@hale.yerkes.uchicago.edu) \and
Space Telescope Science Institute, 3700 San Martin
        Drive, Baltimore, MD 21218  (jking@stsci.edu) \and
Royal Greenwich Observatory, Madingley Road, Cambridge CB3 0EZ, UK
        (sgr@ast.cam.ac.uk)}
 
\date{Received ; accepted }
\offprints{D.K.\ Duncan}

\maketitle

\begin{abstract}

The Goddard High Resolution Spectrograph (GHRS) of the {\it Hubble
Space Telescope (HST)} has been used to observe the boron 2500
\AA\ region of \bd-13.  At a metallicity of [Fe/H]=$-3.00$ this is the
most metal-poor star ever observed for B. Nearly 26 hours of exposure
time resulted in a detection.  Spectrum synthesis using the
latest Kurucz model atmospheres yields an LTE boron abundance of 
log \eps(B)$= +0.01\pm0.20$.  This value is consistent with the linear
relation of slope $\sim$1.0 between log \eps(B$_{\rm LTE}$) and 
[Fe/H] found for
10 halo and disk stars by Duncan \etal\ (1997).  Using the NLTE
correction of Kiselman \& Carlsson (1996), the NLTE boron abundance is
log \eps(B)$= +0.93\pm0.20$.
This is also consistent with the NLTE relation determined by
Duncan \etal\ (1997) where the slope of log \eps(B$_{\rm NLTE}$) vs.
[Fe/H] is $\sim$0.7.

These data support a model in which most production of B and Be comes
from the spallation of energetic C and O nuclei onto protons and He
nuclei, probably in the vicinity of massive supernovae in star-forming
regions, rather than the spallation of cosmic ray protons and alpha
particles onto CNO nuclei in the general interstellar medium.


\end{abstract}

\section{Introduction}

The light elements lithium, beryllium, and boron are of great interest
out of proportion to their very low abundances, having implications
in Big Bang Nucleosynthesis and stellar structure, as well as in
constraints on models of galactic chemical evolution.

The ``canonical'' theory of the origin of the elements Li, Be, and B
was first presented by Reeves, Fowler, \& Hoyle (1970) and further
developed by Meneguzzi,
Audouze, \& Reeves (1971), and then Reeves, Audouze, Fowler, \&
Schramm (1973).
In this model, most light element formation can be accounted for by
galactic cosmic rays (GCR) impinging on the interstellar medium (ISM),
assuming a constant flux of GCRs through the life of the
Galaxy and making reasonable assumptions about CR confinement by the
Galactic magnetic field.  Meneguzzi \etal\ (1971) also
introduced the idea of a large (up to three orders of magnitude)
increase in the low energy (5-40 MeV nucleon$^{-1}$) CR flux; since CRs
in this energy range are mostly shielded from the Solar System by the
solar wind, they are not detectable.  This additional CR flux increased
the production of all light elements and matched the isotopic ratios
and total abundances to the accuracy known at the time.

Reeves \& Meyer (1978) added the additional constraint that models
should match not only present-day abundances but their evolution
throughout the life of the Galaxy.  Their conclusions were similar to
MAR, except that they had to introduce infall of light-element-free
matter into the Galactic disk to match the evolution with time.  In
retrospect, it can be seen that the data they were fitting were sparse
and not very precise.  With the launch of the the {\it Hubble Space
Telescope} ({\it HST}) and the availability of uv-sensitive CCD
detectors (B is usually observed at $\lambda$2500 and Be at
$\lambda$3130), data are now much more numerous and accurate, and
abundances can be traced from the epoch of formation of the Galactic
halo until the present day.

In the past several years, the evolution of Li, Be, and B has been used
as a test of different models of the chemical and dynamical evolution
of the Galaxy.  For example, in the models of Vangioni-Flam \etal\
(1990), Ryan \etal\ (1992), 
and Prantzos \etal\ (PCV; 1993), light element production
depends on the intensity and shape of the GCR spectrum, which in turn
depends on the supernova (SN) and massive star formation rates.  It
also depends on the rise of the (progenitor) CNO abundances and the
decline of the gas mass fraction, which is affected by rates of infall
of fresh (unprocessed) material and outflow, e.g. by SN heating.  Other
things being equal, at early times when target CNO abundances were low,
light element production would be much lower for a given CR flux than
presently, when the ISM abundances are higher.  PCV found that even
with these numerous adjustable parameters, no time-independent CR
spectrum can reproduce the evolution of light element abundances.  By
assuming a very particular form of time variation of the CR flux
(greatly enhanced at early epochs), they
were able to (barely) fit the evolution of the abundances.  

The present investigation supports a different solution to the problem
of the origin of the light elements.  Duncan \etal\ (1997) present B
abundances in a large number of stars ranging in metallicity from 
$\sim$solar
to [Fe/H]~$\sim -2.8$.  They find that (LTE) B follows metals in direct
proportion from the earliest times (very metal-poor stars) to the
present, with little if any change of slope between halo and disk
metallicities.  A straightforward interpretation of this is that the
rate of production of B and Be does {\it not} depend on the CNO
abundances in the ISM, and that the production site is
associated with the production site for metals.  This would be true if
the spallation process most important for light element production is
not primarily protons and $\alpha$ particles colliding with CNO nuclei
in the ISM but rather C and O nuclei colliding with
ambient protons and $\alpha$ particles, probably in regions of massive star
formation (cf. Vangioni-Flam \etal\ 1996 and Ramaty \etal\ 1997).

This paper focuses specifically on the B measurement in \bd-13; it is
consistent with (and was used to help determine) the 
relationships between LTE and NLTE B and [Fe/H] seen in Duncan \etal\ (1997).
The data suggest that B production at the lowest metallicities occurs
in the same way as today, and thus support the new
description of galactic light element production.

It is possible that recent GRO satellite observations of gamma rays
from the Orion Nebula (Bloemen \etal\ 1994; Bykov \& Bloemen 1994)
provide direct evidence of C and O spallation occurring today, even
though not all instruments on GRO detected evidence of such spallation
(Murphy \etal\ 1996).  The light element data alone, however, are the
strongest evidence in support of a new model of their production.

\section{Observations and data reduction} 

The Goddard High-Resolution Spectrograph (GHRS) of the {\it Hubble
Space Telescope} ({\it HST}) was used with the G270M grating to obtain
spectra of resolution 26\,000 (0.025 \AA\ pixel$^{-1}$) in the
$\lambda$2500 \boi\ region of \bd-13.  Data was collected for a total
of 25.39 hours exposure time in six separate visits of the satellite
over several months from December 1994 to May 1995.  The S/N of the
final spectrum was 100 per pixel (200 per resolution element), limited
only by photon statistics.  The spectrum obtained appears in Fig.~1.

These data were combined with that from 
10 other stars, including three previously analyzed
by Duncan, Lambert, \& Lemke (1992) and seven new stars 
([Fe/H] = $-$0.45 to $-$2.75), in a separate
investigation (Duncan \etal\ 1997).  Special care was taken
to analyze each of the stars in a similar manner, and the stars from
Duncan, Lambert, \& Lemke (1992) were re-analyzed to assure
consistency.  See Duncan \etal\ (1997) for more details.

Data reduction followed the standard {\it HST} procedure, using the
IRAF {\sf stsdas} package, combining the quarter-stepped,  FP-SPLIT
data within each visit
with the tasks {\sf poffsets} and {\sf specalign}, which perform a
generalized least-squares solution for the photocathode granularity and
the true spectrum.  
Since {\sf specalign} equally weights the input
spectra, exposures from separate visits were shifted and combined using
the separate tasks {\sf shiftlines} and {\sf scombine}, weighting the
data by the number of counts in the raw spectrum from each visit.  By
comparing subsets of the data, it was confirmed that the photon
statistics limit was reached and that the line in question was present
in both data sets if the data was arbitrarily divided in half, and not
substantially altered by any exposure or by texture on the
photocathode.

\begin{figure}
\resizebox{\hsize}{!}{\includegraphics{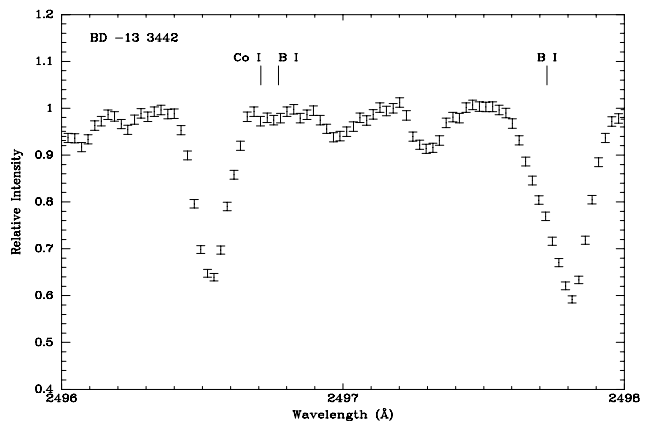}}
\caption{Spectrum of \bd-13. The instrumental resolution is 0.096 \AA, 
and the observational data shown are oversampled by a factor of about 2.}
\end{figure}

\section{Abundance analysis}

Spectrum synthesis using the latest Kurucz model atmospheres was used
to determine log \eps(B), with particular attention paid to the errors.
An accurate determination of the uncertainties is required in the
subsequent discussion of the results, to permit judgement of the
goodness-of-fit of models of Galactic chemical evolution and to
discriminate among the different proposed mechanisms of light element
formation, particularly because this star is the most metal-poor yet
investigated for boron.

\subsection{Spectrum synthesis}

Spectrum synthesis was done using the SYNTHE program distributed by
Kurucz (1993) on CD-ROM.  This program assumes local thermodynamic
equilibrium (LTE) in determining level populations and calculating the
emergent spectrum.  Scripts written by Steve Allen (UC Santa Cruz) were
used to run the program on Unix SparcStations. The most recent grid of
model atmospheres, also released by Kurucz on CD-ROM, includes the
blanketing of almost 60 million lines, both atomic and molecular.  As
in our parallel study of the Galactic evolution of boron, (Duncan
\etal\ 1997), we used the linelist of the Duncan \etal\ (1997b) study of
B in the Hyades giants, which consists almost entirely of
laboratory-measured lines.  It has been carefully tested in order to
fit both the Hyades giants and various metal-poor stars in the B
$\lambda$2500 region.  This list was constructed by starting with
Kurucz's LOWLINES list (CD-ROM \#~1), adjusting (reducing) some
\gf\ values where necessary, and adding some predicted \fei\ lines in
order to achieve a good match to the absorption features near the
\boi\ lines.  The \gf\ value for the \coi\ 2496.708~\AA\ line which
blends with the short wavelength component of the \boi\ doublet is the
same as that of the higher resolution study of Edvardsson
\etal\ (1994).  In our linelist, the \gf\ values and wavelengths of the
\boi\ resonant doublet were selected from O'Brian \& Lawler (1992).
They differ by only 0.03~dex and $<$0.01~\AA\ from the values given by
Kurucz.  Our abundances are derived from fitting the shorter
wavelength, less blended component of the B doublet, with the longer
wavelength component examined as a consistency check.

As in Duncan \etal\ (1997), the Kurucz default abundance of
log \eps(B)=2.57 is used as our solar B value.  This decision
affects the choice of the zero point of the abundance scale and does
not affect the interpretation of trends in the B data discussed in the
present paper.

The Kurucz model grid is given in steps of 250~K in \teff, 0.50 in
\logg, and 0.50~dex in [Fe/H] (at least for halo
stars).  We determined the
physical parameters and their uncertainties for the program star from
an extensive search of the literature and performed calculations at the
nearest grid points.  Within the range of values determined from the
literature, models from the Kurucz grid which best fit our entire uv spectrum
were chosen.  
Experiments with linear interpolation between grid models
were also performed.

The Kurucz models we used assume all metals scale together, whereas it
is known that $\alpha$-elements are enhanced at low metallicities.  In
Duncan \etal\ (1997), testing showed that the effect of
$\alpha$-enhancement on the structure of the model atmosphere produces
very small effects on the calculated B abundances.  In particular, the
three stars originally studied by Duncan, Lambert, \& Lemke (1992) were
analyzed by them using atmospheres with $\alpha$-enhanced composition
from Gustafsson \etal\ (1975).  Analyzing them
with the methods of the present paper produced differences in B
abundance within 0.04~dex, much lower than the other errors
associated with the abundance determination.  The direct effect of an
enhanced Co abundance could be larger due to the particular line
\coi\ $2496.708$ \AA\ which is blended with the shorter wavelength B
component.  This is discussed further in the section on uncertainties.

The atmospheric models we used were computed with a microturbulent
velocity of 2~km~s$^{-1}$. We did not find it necessary to change to the
commonly accepted value of 1.5~km~s$^{-1}$ for halo stars (Magain 1989),
because analysis of several curves of growth showed that the dependence
of B abundances on this stellar parameter is negligible,
though microturbulence would affect the blending Co line, which we 
discuss more fully below.

Fig.~2 presents spectrum synthesis fits for \bd-13; the fits
presented are calculated to have log \eps(B)=+0.3, +0.1, and $-$0.4,
the latter indicates what the abundance would be if this star
had the solar B to metals ratio.  Table~1 presents
the stellar parameters and B abundance for \bd-13. 

Table~1 also presents parameters from the three most metal-poor stars
of our recent study 
(Duncan \etal\ 1997): 
HD~140283, which for several years was the most metal-poor star 
with detected B (Duncan \etal\ 1992), and BD~$+$3~740.
The metallicities determined by us for these stars are consistent 
with the ranges found by other workers (see compilation by Cayrel de 
Strobel \etal\ 1997).  
Fig.~3 compares the observed spectra of \bd-13\ with that of 
HD~140283 and BD~$+$3~740.
It is clear that \bd-13\ is the most metal-poor of the three stars.

\begin{table}
\caption{Stellar parameters and B abundances}
\begin{flushleft}
\begin{tabular}{cccccc} \hline
star & \teff & \logg & [Fe/H] & log$\epsilon(B)_{LTE}$ & 
	log$\epsilon(B)_{NLTE}$ \\ 
\hline
\bd-13 & 6250  & 3.75 & $-$3.00 & 0.01 & 0.93 \\
BD~+3~740 & 6125 & 3.5 & $-$2.75 & 0.22 & 1.04 \\
HD~140283 & 5640 & 3.5 & $-$2.60 & $-$0.10 & 0.35 \\
\hline
\end{tabular}
\end{flushleft}
\end{table}

\begin{figure}
\resizebox{\hsize}{!}{\includegraphics{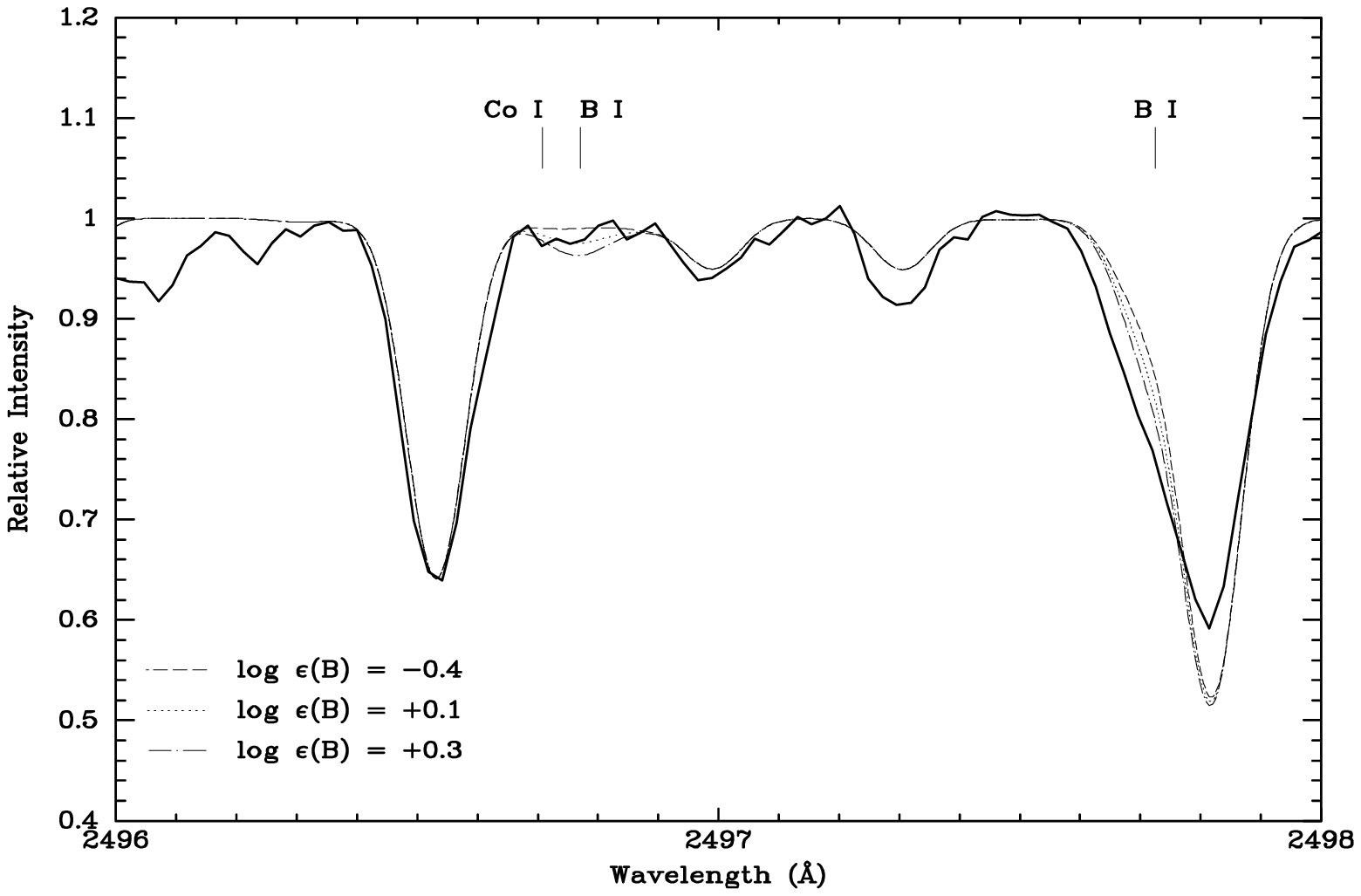}}
\resizebox{\hsize}{!}{\includegraphics{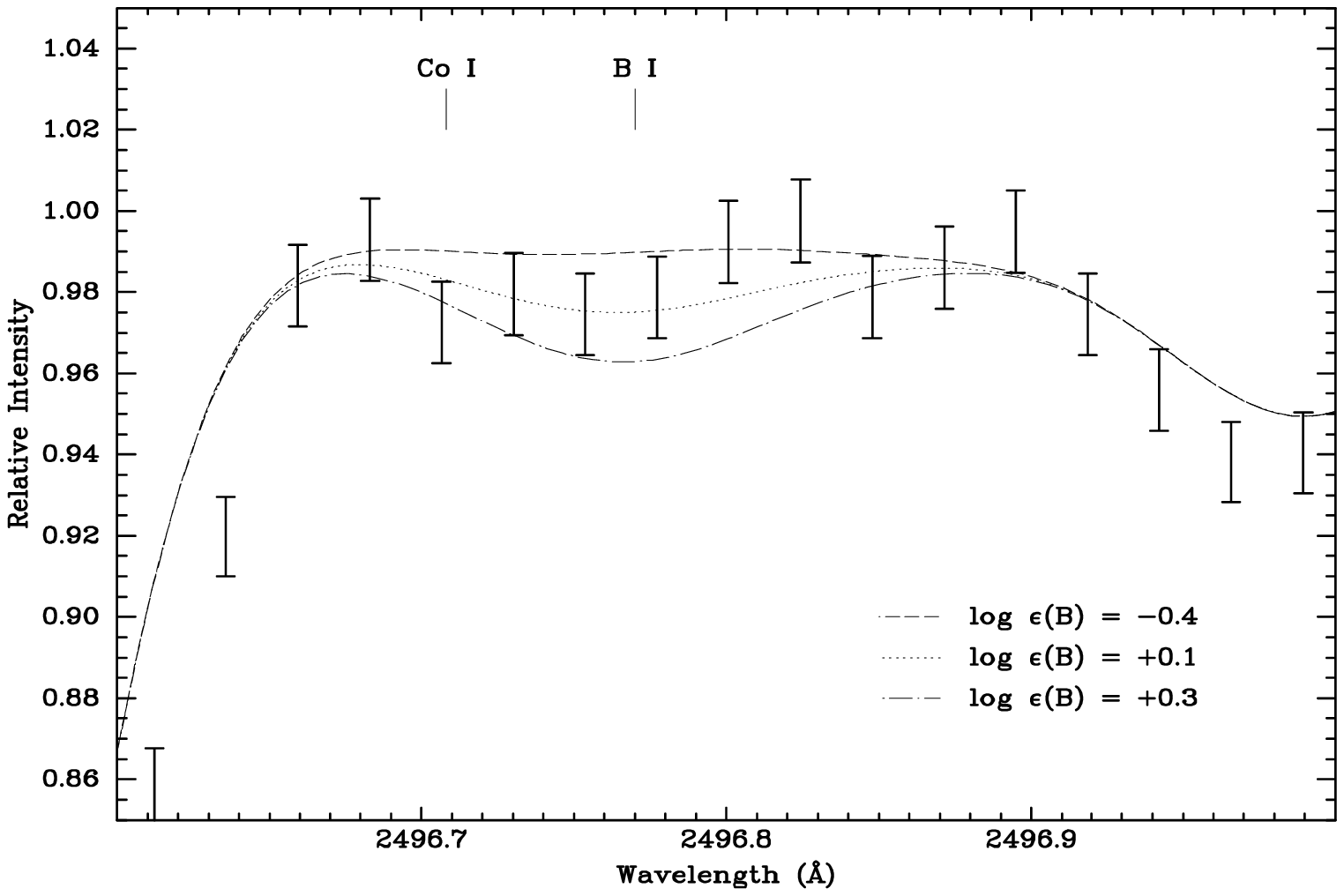}}
\caption{{\bf a} Spectrum synthesis fits for \bd-13. {\bf b} Enlarged
view of fits near the B lines.}
\end{figure}

\begin{figure}
\resizebox{\hsize}{!}{\includegraphics{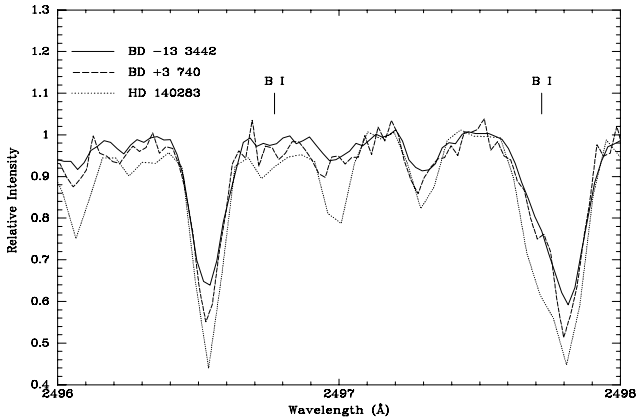}}
\caption{Comparison of \bd-13, HD~140283, and BD~$+$3~740.}
\end{figure}

\subsection{Uncertainties}

Considerable care was taken in determining the error of the abundance
determination. Sources of error we considered included uncertainties in
stellar \teff, \logg, and [Fe/H] (all of which change the atmosphere
used in the analysis), continuum placement, photon statistics in the
points defining the line itself, and the effect of metal lines which
blend with the B features.

In such a metal-poor star the continuum is relatively easy to define
but the B line is weak.  Table~2 presents, in order, effects due to
uncertainties in stellar \teff, \logg, [Fe/H], continuum location and
photon statistics in the B line, as well as the total (net) error.  We
quantified the dependence of the B abundance on changes in the main
stellar parameters by running multiple syntheses and curves of growth,
although curves of growth are less reliable since the B lines are
blended. From available literature determinations for this
star, we considered $\pm$75~K, $\pm$0.30, and $\pm$0.10~dex to be 
typical uncertainties to be associated with \teff, \logg, and [Fe/H]
respectively.
Errors for BD~+3~740 and HD~140283 are also included in Table~2
for comparison.

Uncertainty in stellar metallicity in metal-poor stars causes error due
to lines which blend with the B features.  At our resolution, the
\coi\ 2496.708 \AA\ feature blends with \boi\ 2496.772 \AA.  Assumption
of a higher metallicity attributes more of the blend to Co, decreasing
the derived B abundance, and vice-versa for a lower assumed
metallicity.  The column ``[Fe/H]'' in Table~2 gives the amount of this
direct effect when the metallicity is uncertain by $\pm$0.1 dex.

McWilliam \etal\ (1995) studied Co in their spectroscopic
analysis of the most metal-poor stars.  They present evidence that Co
remains scaled-solar to [Fe/H] $\sim -$2.5, but that it
increases by $\sim$0.5 dex as [Fe/H] decreases to $-$3.0.  Since our
spectra do not resolve the Co-B blend, overabundance of Co means that
the B abundances should be decreased.  However, the blend between the
Co and B features is only partial and an increase in Co of a given amount
produces a decrease in B abundance of a smaller amount. 
Synthesis testing with [Co/H]
increased by 0.1 and 0.3 dex caused a decrease in B abundance of
about 0.05 and 0.15 dex respectively.  The data are not consistent with 
[Co/H] enhanced by 0.5 dex.  Since we have no way
of deblending the lines, this error is not included in Table~2;
it should be considered a systematic
effect which could affect the derived abundance, and at worst case,
our abundance represents an upper limit.  
That this effect is unlikely to be large is supported by the good
agreement of the B abundance derived by Duncan \etal\ (1997) for the
slightly more metal-rich star HD~140283 with that of the higher
resolution study of Edvardsson \etal\ (1993), which partially resolves
the Co-B blend.  The spectrum for \bd-13 is similar in resolution to
that for HD~140283 in Duncan \etal\ (1997).

\begin{table}
\caption{Uncertainties affecting the derived B abundance}
\begin{flushleft}
\begin{minipage}{3in}
\begin{tabular}{ccccccc} \hline
star & \teff & \logg & [Fe/H] & Continuum & Photon & {\bf Net}  \\
    &$\pm$75~K & $\pm$0.30 & $\pm$0.10 & $\pm$0.5\% & Statistics& {\bf Error}\\ 
\hline
\bd-13 & $\pm$0.09 & $\pm$0.03 & $\pm$0.10 & $\pm$0.10 & $\pm$0.10 & 
	{\bf $\pm$0.20} \\
BD~+3~740\footnote{Final boron abundance 0.22 LTE and 1.04 NLTE.} & 
	$\pm$0.08 & $\pm$0.05 & $\pm$0.10 & 
	$\pm$0.10\footnote{$\pm$1.5\% in the continuum.} & $\pm$0.23 
	& {\bf $\pm$0.29} \\
HD~140283\footnote{Final boron abundance $-$0.10 LTE and 0.35 NLTE.} & 
	$\pm$0.08 & $\pm$0.05 & $\pm$0.10 & 
	$\pm$0.10\footnote{$\pm$1.5\% in the continuum.} & $\pm$0.06 & 
	{\bf $\pm$0.18} \\
\hline
\end{tabular}
\end{minipage}
\end{flushleft}
\end{table}

\subsection{NLTE abundances}

The largest source of systematic errors in our B abundance is likely to
be due to NLTE effects.  As calculated by Kiselman (1994), and Kiselman
\& Carlsson (1996), NLTE effects increase B abundances most
significantly in the most metal-poor stars.  The relatively hot uv
radiation fields in metal-poor stars drive the detailed balance of the
B resonance lines away from statistical equilibrium.  Overionization
and optical pumping tend to weaken the \boi\ 2498 \AA\ doublet, which
leads to an underestimate of the B abundance.  NLTE calculations are
difficult, being limited both by the incomplete understanding of the
atomic physics involved and especially by the uncertainties in the uv
flux values at the precise absorption wavelengths.  The Kiselman \&
Carlsson (1996) study is improved over the Kiselman (1994) one in that
it models the effects of large numbers of lines in the background
radiation field, which tend to decrease NLTE effects.   Results of the
two studies differ by approximately 0.1 dex; present NLTE calculations
are necessarily less accurate than that amount.  The Kiselman \&
Carlsson NLTE correction for \bd-13 is $+0.92$ dex. 

\section{Discussion}

Fig.~4 shows the 
LTE abundances as a function of [Fe/H] from
stars analyzed in the larger investigation of Duncan
\etal\ (1997), with the point for \bd-13 from the present investigation
emphasized.  It can be seen that there is an approximately linear
relation between $\log$ \eps(B$_{\rm LTE}$) and [Fe/H] over both disk
and halo metallicities, and that \bd-13\ is consistent with this
relationship.  A least-squares fit to all the data of Fig.~4 (allowing
for errors in both coordinates) yields a slope of 0.96$\pm$0.07 and a
reduced chi square, $\chi_{\nu}^2$, of 0.71, indicating an excellent
fit.  If NLTE abundances are used, as shown in Fig.~5, the slope is
0.70$\pm$0.07 and $\chi_{\nu}^2 = 1.63$.  Although it is true that
\bd-13\ is used to determine this line, it is important to note that
this star is quite consistent with the trend defined by the other
stars.  Inspection and $\chi_{\nu}^2$ tests confirm this.

\begin{figure}
\resizebox{\hsize}{!}{\includegraphics{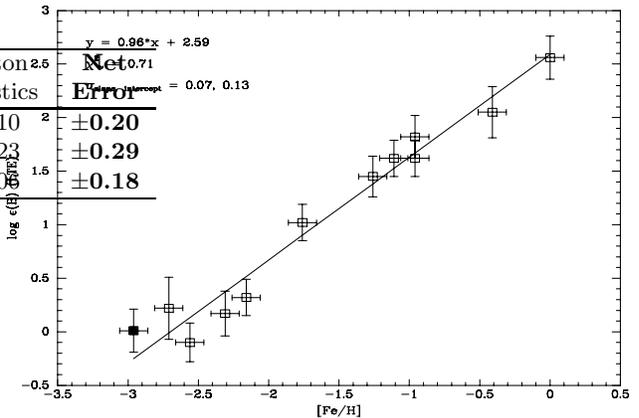}}
\caption{LTE B abundances from Duncan \etal\ (1997) with the program
star highlighted.}
\end{figure}
 
\begin{figure}
\resizebox{\hsize}{!}{\includegraphics{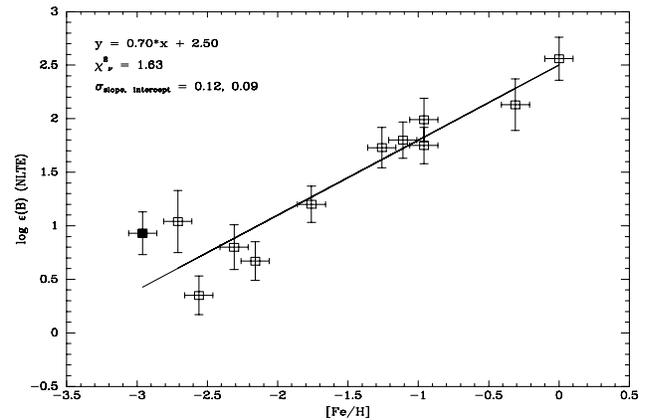}}
\caption{NLTE B abundances from Duncan \etal\ (1997) with the program
star highlighted.}
\end{figure}

\subsection{Comparison to standard models and a new model for light 
element production}

The slope of close to 1 suggesting a primary process is {\it not}
expected from canonical models of CR spallation in the ISM, which
predict a secondary process and thus a steeper relation.  In a
secondary process the rate of light element production depends on the
product of the abundance of target CNO nuclei and the CR flux, both of
which vary with time.  If SNe are the source of the target nuclei and
the ISM is well-mixed, the ISM metallicity is proportional to the integral
(total) number of SN up to a given time.   If, as is commonly supposed,
SNe also seed the acceleration mechanism which produces CRs, the CR
flux is proportional to the SN rate.  The result is light element
abundances which vary quadratically with the metallicity of the ISM, or
a logarithmic slope of 2 (Prantzos \etal\ 1993).
Figs.~4 and 5 show that such a slope is certainly not consistent with
our data.
Duncan \etal\ (1997) discuss these issues in greater detail.

As was discussed by Duncan, Lambert, and Lemke (1992), the data for the
first three metal-poor stars observed for B already seemed to show a
linear (in the log) relationship with [Fe/H], suggesting some primary
production mechanism rather than the secondary mechanism in the ISM
described above. This idea has been modelled in detail by Cass\'e
\etal\ (1995), Ramaty \etal\ (1995 and 1997), Lemoine \etal\ (1997),
and Vangioni-Flam \etal\ (1996).  In the new scenario, B and Be are
primarily produced by the spallation of C and O onto protons and
$\alpha$ particles.  Such a process could occur near massive star SNe,
where the particle flux would be very non-solar in composition;
depleted in H and He and especially enriched in O and C.  Vangioni-Flam
\etal\ find that a composition matching either winds from massive
(Wolf-Rayet; WR) stars in star-forming regions or massive star SNe
produce a flux of O and C which, after further acceleration, can
reproduce both the magnitude and slope of B production seen in Figs.~4
and 5 through collisions with protons and $\alpha$ particles.  As
Ramaty \etal\ point out, production of some additional $^{11}$B by the
neutrino process (Woosley \etal\ 1990) is not ruled out, and may be
favored on energetic grounds.  Nevertheless the bulk of the B and Be
would be produced from the spallation process.

Although the NLTE correction to the B abundance of \bd-13 is relatively
large and tends to raise the B abundance above the line in Fig.~5, two
other effects not included here would tend to move it closer to the
curve.  One is the effect of the blending Co line discussed above,
which could reduce the B abundance as much as $\sim$0.15 dex.  The
other is the fact that if the spallation producing light elements is
caused by O (and to a lesser extent C), $\log$ \eps(B) should be
plotted against [O/H] rather than [Fe/H].  As is well-known, very
metal-poor stars are overabundant in O compared to Fe (moving points
representing the most metal-poor stars to the right in a figure with O
on the x-axis).  Duncan \etal\ (1997) make such a plot, and demonstrate
that although the measurement errors in O are greater than those for
Fe, when all the metal-poor stars are considered together a straight
line of slope 1.10$\pm$0.14 fits the LTE abundances, and one of slope
0.82$\pm$0.10 the NLTE abundances.  
However, oxygen abundance measurements are also 
surrounded by greater uncertainty than are iron measurements, and 
systematic errors in the oxygen abundance which depend on metallicity will
affect the derived slope.

\section{Summary}

The present investigation has presented an analysis of B in \bd-13.  At
a metallicity of [Fe/H]=$-3.0$, this is the most metal-poor star ever
observed for B.  Careful spectrum synthesis was performed to determine
both the B abundance and its uncertainty. NLTE corrections were also
considered.

We find that \bd-13\ is consistent with the straight line of slope
close to 1 between log \eps(B) vs.\ [Fe/H] or log \eps(B) vs.\ [O/H]
found by Duncan \etal\ (1997).  This suggests that the production of
light elements such as B and Be is directly related to the production
of heavier elements.  Canonical models which only include spallation
by the process of protons and $\alpha$ particles impinging 
onto the general Galactic ISM predict
a steeper relation which fails to fit all the data now available.

Spallation via the reverse process of energetic C and O nuclei onto
protons and $\alpha$ particles does much better
in explaining the linear relationship observed.  This reverse process
was historically discarded as a minor effect (e.g. 20\% according to
the calculations of Walker \etal\ 1993) since the canonical theory
assumed a CR and ISM composition similar to that in the solar system.  A
population of low-energy CRs
enriched in C and O and depleted in H and He can produce light
elements by spallation onto the ambient medium.  The
metallicity of the medium would have little effect, and the process
would operate the same throughout the life of the Galaxy, leading to
the observed slope of 1.  A model in which most light element
production comes from CR spallation of C and O nuclei onto protons and
$\alpha$ particles in the vicinity of massive supernovae in
star-forming regions fits the data well.  If particle compositions are
taken from calculations of either massive star SNe or WR stars,
spallation can reproduce both the magnitude and slope of the observed B
evolution (Ramaty \etal\ 1996) while maintaining consistency with the
other light element abundances. 
The observations of B in \bd-13 provide
the strongest evidence of how CR spallation operated in the very early
galaxy.

\begin{acknowledgements}
We thank Michel Cass\'e, Reuven Ramaty, Hubert Reeves, Elizabeth
Vangioni-Flam, and Stan Woosley for their comments.

This research was based on observations obtained with the NASA/ESA {\it
Hubble Space Telescope} through the Space Telescope Science Institute,
which is operated by the Association of Universities for Research in
Astronomy, Inc., under NASA contract NAS5-26555.  This research has
made use of NASA's Astrophysics Data System Abstract Service.

CPD gratefully acknowledges support provided by the University of Hawaii 
Foundation, and subsequently by NASA through grant HF-1042.01-93A from the
Space Telescope Science Institute.

\end{acknowledgements}

\end{document}